\pdfoutput=1
\documentclass[11pt]{article}

\usepackage{graphicx}
\usepackage{amsmath}
\usepackage{dcolumn}
\usepackage{bm}
\usepackage{slashed}
\usepackage[utf8]{inputenc}
\usepackage{authblk}

\newcommand{\be}{\begin{equation}}
\newcommand{\ee}{\end{equation}}
\newcommand{\ba}{\begin{eqnarray}}
\newcommand{\ea}{\end{eqnarray}}
\newcommand{\no}{\nonumber \\}
\newcommand{\gsim}{\mathrel{\hbox{\rlap{\lower.55ex \hbox {$\sim$}}
                   \kern-.3em \raise.4ex \hbox{$>$}}}}
\newcommand{\lsim}{\mathrel{\hbox{\rlap{\lower.55ex \hbox {$\sim$}}
                   \kern-.3em \raise.4ex \hbox{$<$}}}}

\def\roughly#1{\mathrel{\raise.3ex\hbox{$#1$\kern-.75em%
\lower1ex\hbox{$\sim$}}}}
\def\lsim{\roughly<}
\def\gsim{\roughly>}

\def\({\left(}
\def\){\right)}
\def\[{\left[}
\def\]{\right]}
\def\<{\langle}
\def\>{\rangle}
\def\pd{\partial}

\def\l{{\lambda}}

\def\d{{\delta}}
\def\D{{\Delta}}

\def\O{{\Omega}}
\def\e{{\epsilon}}

\def\a{{\alpha}}

\def\g{{\gamma}}

\def\p{{\pi}}

\def\m{{\mu}}
\def\n{{\nu}}

\def\s{{\sigma}}

\def\t{{\tau}}

\def\h{{\eta}}

\newcommand{\MeV}{\text{MeV}}
\newcommand{\mx}{\text{max}}
\newcommand{\mn}{\text{min}}

\title{\bf Entropy Production in Dissipationless Hydrodynamics with an order parameter}

\author[1]{Shu Lin}
\author[1]{Gezheng Zhou}
\affil[1]{School of Physics and Astronomy, Sun Yat-Sen University, Zhuhai, 519082, China}

\date{\today}

\begin{document}

\maketitle

\begin{abstract}

We study hydrodynamics coupled to order parameter based on linear sigma model. We obtain numerical solutions for both boost invariant and non-boost invariant solutions. Both solutions show the order parameter rises with oscillations, which persist at late time. The temperature drops with correlated oscillations, which can be approximated by a power law at mid-rapidity. We also find the entropy is conserved in the boost invariant case, but entropy production is seen in non-boost invariant solution. We interpret the entropy production as due to smoothening of inhomogeneity in the off-equilibrium state.

\end{abstract}

\newpage

\section{Introduction}

It is remarkable that relativistic hydrodynamics provides accurate description of bulk evolution of matter produced in heavy ion collisions, which consists of around a thousand particles \cite{Teaney:2001av,Kolb:2003dz}. With the assumption of local equilibrium, the problem of complicated many particle dynamics is reduced to the conservation of energy, momentum and baryon number as equation of motion for relativistic hydrodynamics. Over the past decade, the framework of relativistic hydrodynamics has been furnished in many aspects for phenomenological application in heavy ion collisions: the inclusion of viscous correction leads to more accurate description of the bulk evolution \cite{Teaney:2003kp,Baier:2006gy,Romatschke:2007mq}; the inclusion of noises allows for systematic treatment of fluctuations \cite{Kapusta:2011gt,Akamatsu:2016llw,An:2019osr}; the inclusion of particle momentum anisotropy extends the regime of applicability to earlier time \cite{Florkowski:2010cf,Martinez:2010sc} etc.

Recently the beam energy scan (BES) program in relativistic heavy ion collider (RHIC) is devoted to pinning down possible existence of critical point in the phase diagram of quantum chromodynamics (QCD) \cite{Luo:2012kja}. Hydrodynamic studies of system evolution close to the critical point necessitates the inclusion of critical mode into the present framework of hydrodynamics. This has been pursued both by different group \cite{Nahrgang:2011mg,Mishustin:1998yc,Stephanov:2017ghc,Rajagopal:2019xwg}. The coupling of critical mode and the hydrodynamic modes is found to alter the bulk evolution, in particular near the phase transition, see \cite{Bluhm:2020mpc} for a recent review.

A key quantity in heavy ion collisions is the entropy production. It is known that in complicated dynamics of fireball evolution, entropy is produced in different stages of the collisions, see \cite{Fries:2009wh} and references therein. A significant amount of entropy is produced during hydrodynamic evolution due to dissipation of the hydrodynamic mode. In the presence of critical mode, the dissipation of critical mode can also lead to entropy production \cite{Herold:2018ptm,Kittiratpattana:2019knl}. The goal of this paper is to study entropy production when dissipation of either mode is absent. We will show distinct features of boost invariant solution and non-boost invariant solution, with the former preserving entropy and the latter not.

The paper is organized as follows: In Section 2, we describe the hydrodynamics with critical mode based on linear sigma model. In Section 3, we present numerical results for boost invariant and non-boost invariant solutions and discuss physical implications. We give conclusion and outlook in Section 4.

\section{Hydrodynamics with an order parameter}

We start with the Lagrangian of linear sigma model \cite{Schaefer:2006ds}:
\begin{align}\label{qm_Lag}
{\cal L}={\bar q}\(i{\slashed \pd}-g(\s+i\g_5{\vec \t}{\vec \p})\)q+\frac{1}{2}\((\pd\s)^2+(\pd{\vec \p})^2\)-U(\s,{\vec \p}),
\end{align}
with $q$, $\s$ and ${\vec \p}$ being the quark, sigma and pions fields respectively. The condensation of $\s$ gives mass to quarks $M_q=g\<\s\>$, breaking chiral symmetry. The symmetry is broken through the potential $U$ given by
\begin{align}\label{U}
U(\s,{\vec\p})=\frac{\l}{4}\(\s^2+{\vec\p}^2-v^2\)^2-c\s.
\end{align}
Throughout the paper, we use mean-field approximation for $\s$ and ${\vec\p}$. In the absence of isospin chemical potential, pions do not condense thus $\<{\vec\p}\>=0$. $\<\s\>$ is the only order parameter in the model. This is the critical mode to be included in hydrodynamics.
The parameters in \eqref{qm_Lag} and \eqref{U} are fixed as
\begin{align}\label{parameters}
g=M_q/f_\p,\quad c=M_\p^2f_\p,\quad \l=\frac{1}{2f_\p^2}\(M_\s^2-M_\p^2\),\quad v^2=f_\p^2-M_\p^2/\l,
\end{align}
with the experimental input $M_\p=138\MeV$, $M_\s=600\MeV$, $f_\p=93\MeV$.
The condensate $\<\s\>$ is determined dynamically by minimizing the thermodynamic potential $\O=U+\O_{q{\bar q}}$, with the quark contribution $\O_{q{\bar q}}$ given by
\begin{align}\label{Oqq}
\O_{q{\bar q}}(T,\m)=\n_qT\int\frac{d^3k}{(2\p)^3}\big[\ln\(1-n_q(T,\m,k)\)+\ln\(1-n_{\bar q}(T,\m,k)\)\big].
\end{align}
Here $\n_q=2N_cN_f=12$ counts the spins, colors and flavors of the quark field. $n_q$ and $n_{\bar q}$ are Fermi-Dirac distributions for quark and anti-quark respectively:
\begin{align}
n_q(T,\m,k)=\frac{1}{e^{\(\sqrt{k^2+M_q^2}-\m\)/T}+1},\quad n_{\bar q}(T,\m,k)=n_{q}(T,-\m,k).
\end{align}

Following \cite{Nahrgang:2011mg,Mishustin:1998yc}, we treat quark fields as hydrodynamic degree of freedom, which is coupled to sigma mean field. The equations of motion (EOM) for $\s$ and hydrodynamic fields are given by:
\begin{align}\label{eoms}
&D_\m D^\m\s+\frac{\d\O}{\d\s}=0,\\
&D_\m\(T_q^{\m\n}+T_\s^{\m\n}\)=0.
\end{align}
For simplicity, we choose to work with vanishing quark number, so there is no additional equation for quark number conservation and $\m_B=0$.
Note that we choose to write the EOM in curved spacetime with $D_\m$ denoting covariant derivative. This will be useful for adapting to Milner coordinates in the next section. $T_q^{\m\n}=(\e+p)u^\m u^\n-pg^{\m\n}$ and $T_\s^{\m\n}=\pd^\m\s\pd^\n\s-g^{\m\n}\(\frac{1}{2}(\pd\s)^2-U(\s)\)$ correspond to contributions to stress energy tensor from quark and sigma fields respectively.
Using \eqref{eoms}, we can express the divergence of $T_\s^{\m\n}$ field as
\begin{align}
D_\m T_\s^{\m\n}=-\frac{\d \O_{q{\bar q}}}{\d\s}\pd^\n\s.
\end{align}
We can thus rewrite conservation of stress energy tensor as
\begin{align}\label{eomq}
D_\m T_q^{\m\n}=\frac{\d \O_{q{\bar q}}}{\d\s}\pd^\n\s.
\end{align}
This has clear interpretation that the stress tensor from the quark field is conserved up to work and force by the sigma field. We will solve \eqref{eoms} and \eqref{eomq} numerically with the hydrodynamic fields $T(x)$, $u^\m(x)$ and mean field $\s$ in the next section.

Before presenting numerical results, it is instructive to see the implication of \eqref{eoms} and \eqref{eomq} for entropy production. To this end, we define entropy at $\m=0$ from the following form
\begin{align}\label{dp}
dp=sdT+\frac{\pd p}{\pd \s}d\s\;\;\Rightarrow s=\frac{\pd p}{\pd T}\vert_\s.
\end{align}
To derive the evolution of $s$, we contract \eqref{eomq} with $u_\n$ and use $u^2=1$ to obtain
\begin{align}\label{contracted}
\(\e+p\)D_\m u^\m+u^\m\pd_\m\e+\(\e+p\)u^\m u^\n D_\m u_\n=\frac{\d\O_{q{\bar q}}}{\d\s}u_\n\pd^\n\s.
\end{align}
The last term on the left hand side can be dropped by using $u^\n D_\m u_\n=0$. Using $\e+p=Ts$ and \eqref{dp}, we obtain
\begin{align}\label{de}
d\e=Tds-\frac{\pd p}{\pd \s}d\s.
\end{align}
Plugging \eqref{de} into \eqref{contracted}, we obtain
\begin{align}\label{Ds}
D_\m \(su^\m\)=0.
\end{align}
This is the conservation of entropy current. Note that the $d\s$ terms in \eqref{dp} and \eqref{de} corresponding to change of quark mass cancel each other in $s$. In the absence of order parameter, the existence of entropy current in ideal hydrodynamics is expected from the absence of dissipation in the EOM. \eqref{Ds} shows the presence of order parameter does not affect the existence of entropy current with our definition of entropy density.

\section{Solutions with and without boost invariance}

\begin{figure}
\includegraphics[width=0.5\textwidth]{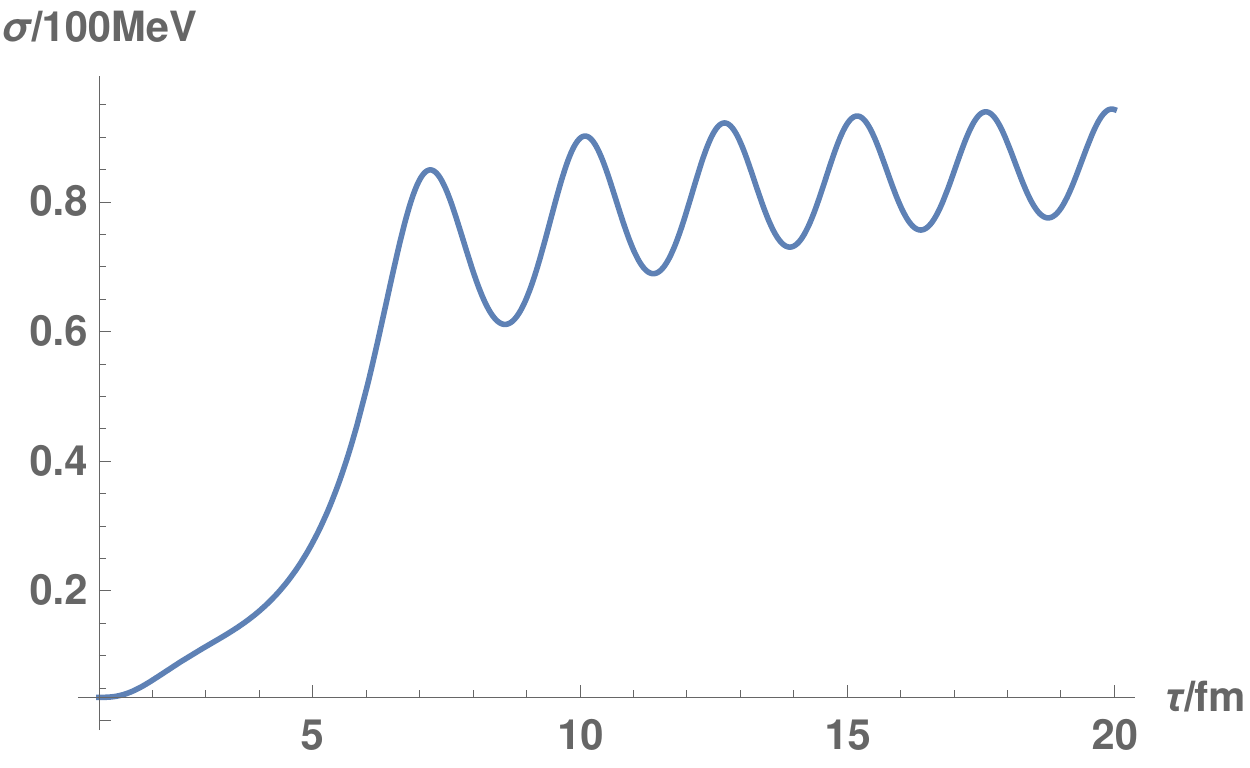}
\includegraphics[width=0.5\textwidth]{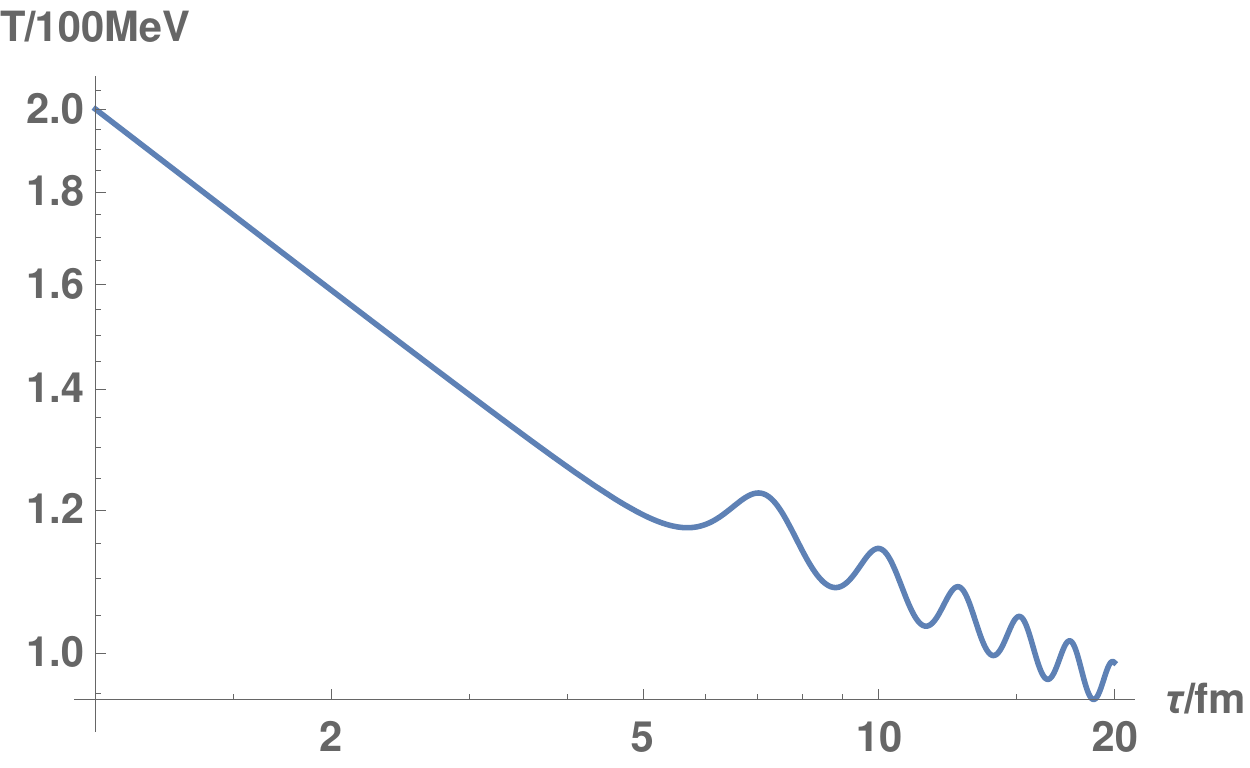}
\caption{\label{Bj200}left panel: $\s$ as a function of $\t$. $\s$ rises with oscillations. The oscillations persist at very late time. right panel: $T$ as a function of $\t$. It shows an approximate scaling law $T\sim\t^{-\a}$. The plots correspond to $T_0=200\MeV$}
\end{figure}

\begin{figure}
\includegraphics[width=0.5\textwidth]{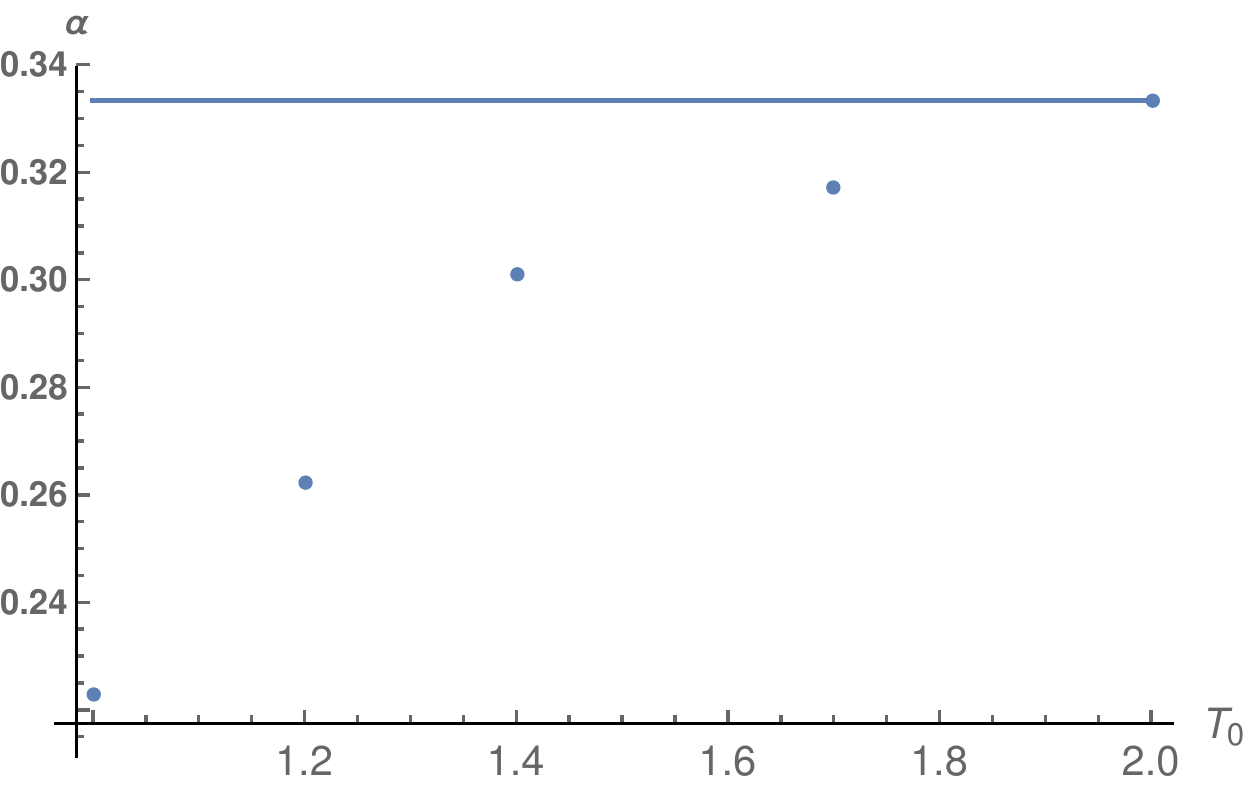}
\caption{\label{alphaBj}$\a$ as a function of $T_0$. It indicates quicker dropping for cases with higher initial temperature. In the high temperature limit, the exponent approaches $c_s^2=1/3$, shown in the same plot.}
\end{figure}

In this section, we solve \eqref{eoms} and \eqref{eomq} numerically. For comparison, we first study solutions with boost invariance, and then solutions without. In both cases, homogeneity in transverse plane is assumed. In terms of proper time $\t=\sqrt{t^2-z^2}$ and spacetime rapidity $\h=\tanh^{-1}\frac{z}{t}$, the explicit form of EOM are given by
\begin{align}\label{eom_exp}
&\pd_\t^2\s+\frac{1}{\t}\pd_\t\s-\frac{1}{\t^2}\pd_\h^2\s+\frac{\d\O}{\d\s}=0,\\
&\pd_\t T_q^{\t\t}+\frac{1}{\t}T_q^{\t\t}+\pd_\h T_q^{\t\h}+\t T_q^{\h\h}-\frac{\d\O_{q{\bar q}}}{\d\s}\pd_\t\s=0,\\
&\pd_\t T_q^{\t\h}+\frac{1}{\t}\pd_\h T_q^{\t\h}+\pd_\h T_q^{\h\h}+\frac{2}{\t}T_q^{\t\h}+\frac{1}{\t^2}\frac{\d\O_{q{\bar q}}}{\d\s}\pd_\h\s=0.
\end{align}
In the boost invariant case, the flow is fixed as $u^\m=(1,0,0,0)$, and the last equation in \eqref{eom_exp} is trivially satisfied with $T$ and $\s$ being functions of $\t$ only. To solve \eqref{eom_exp}, we use the following initial conditions:
\begin{align}\label{ic}
T(\t=\t_0)=T_0,\quad \s(\t=\t_0)=\s_0,\quad \pd_\t\s(\t=\t_0)=0.
\end{align}
We fix $\s_0$ to be the equilibrium value of $\s$ at $T_0$.
In Fig.~\ref{Bj200}, we show the $\t$ dependencies of $T$ and $\s$. As temperature drops due to the expansion, the sigma field rises in an oscillatory fashion. The oscillation persists at very late time in the absence of dissipation. Similar behavior is also observed in \cite{Herold:2018ptm}. Since the dynamics of $\s$ is coupled with $T$, oscillation is also present in $T(\t)$. Fig.~\ref{Bj200} also indicates approximate scaling for temperature $\s$: $T\sim \t^{-\a}$. We extract the exponent $\a$ as a function of the initial temperature $T_0$. In Fig.~\ref{alphaBj}, we show the $T_0$ dependence of $\a$. $\a(T_0)$ is an increasing function, indicating that the dropping of the temperature is faster for higher initial temperature.
In the high temperature limit, $\a\to\frac{1}{3}$. This is an expected limit, as chiral symmetry is essentially restored with $\s\simeq0$. The scaling of the temperature is given by Bjorken solution with $T\sim \t^{-c_s^2}$ and $c_s^2=\frac{\pd p}{\pd\e}\simeq\frac{1}{3}$. 
The total entropy in the boost invariant case can be shown to be strictly conserved. To see that, we write out the explicit form of \eqref{Ds} as
\begin{align}\label{Ds_Bjorken}
\pd_\t s+\frac{1}{\t}s=0,\Rightarrow \frac{1}{\t}\pd_\t\(\t s\)=0.
\end{align}
Note that $\t s d\h d^2x_\perp$ is the entropy per unit volume in Milner coordinates. It follows that entropy is strictly conserved. We also verify this numerically from the definition of entropy \eqref{dp}.

\begin{figure}
\includegraphics[width=0.5\textwidth]{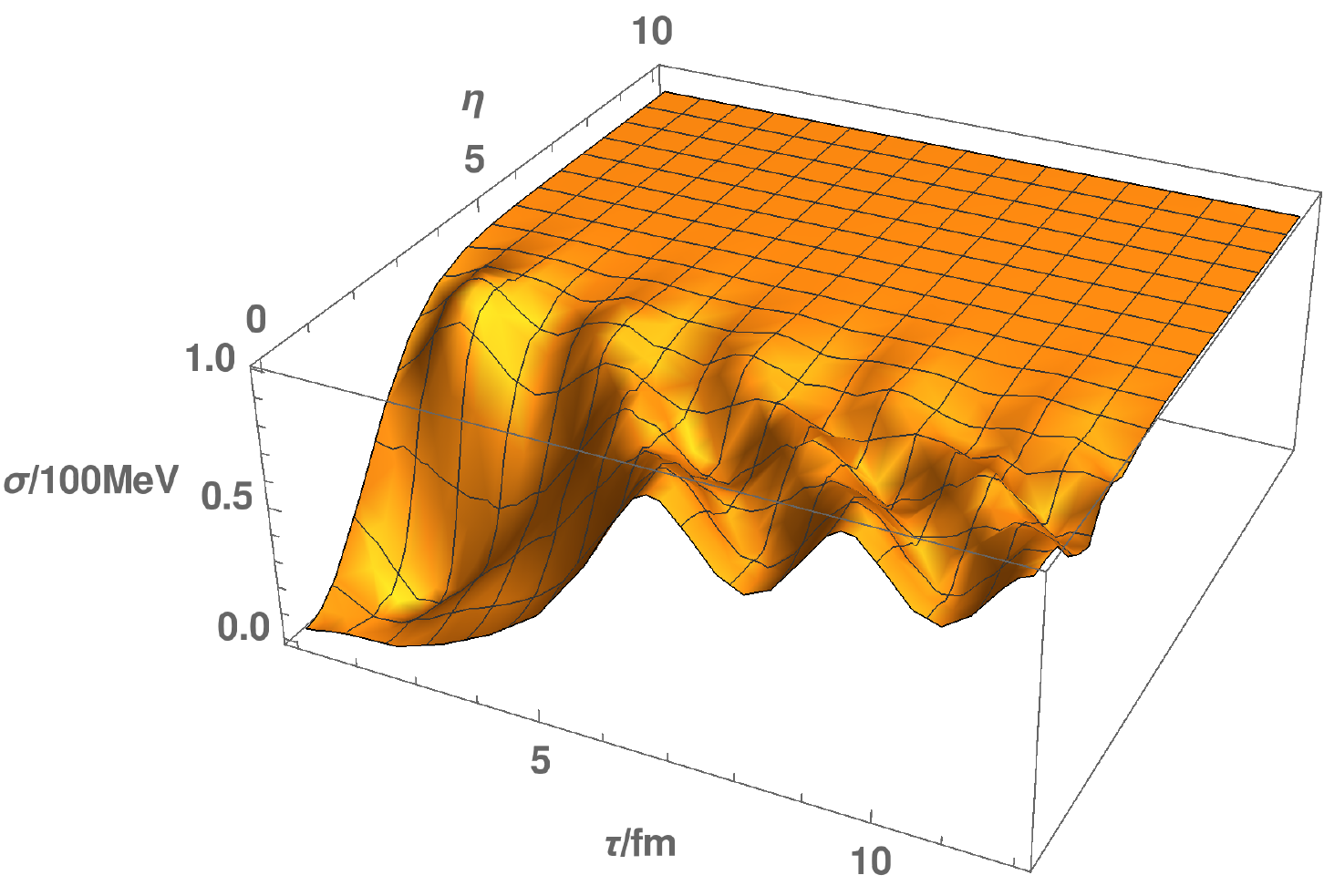}
\includegraphics[width=0.5\textwidth]{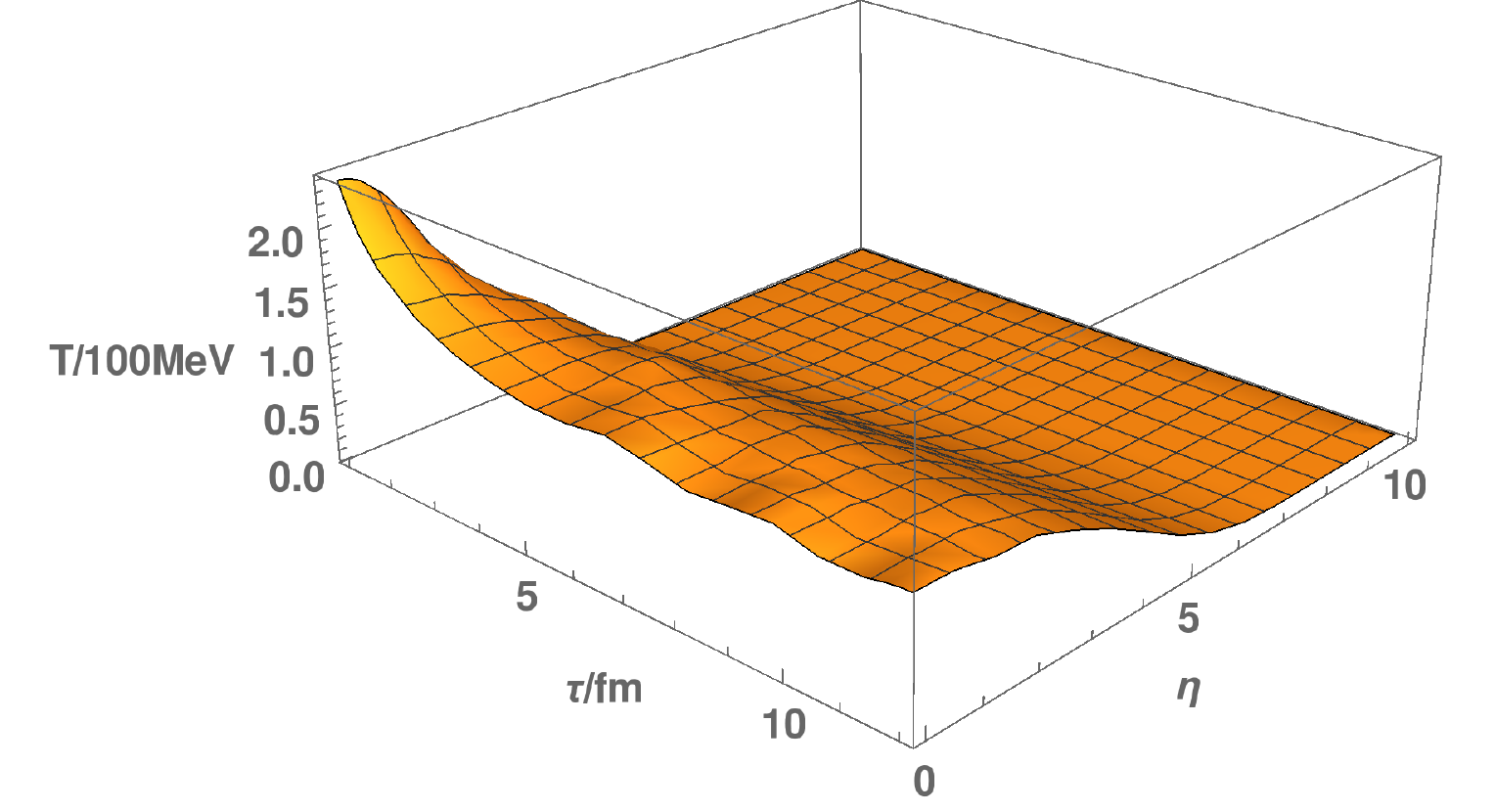}
\includegraphics[width=0.5\textwidth]{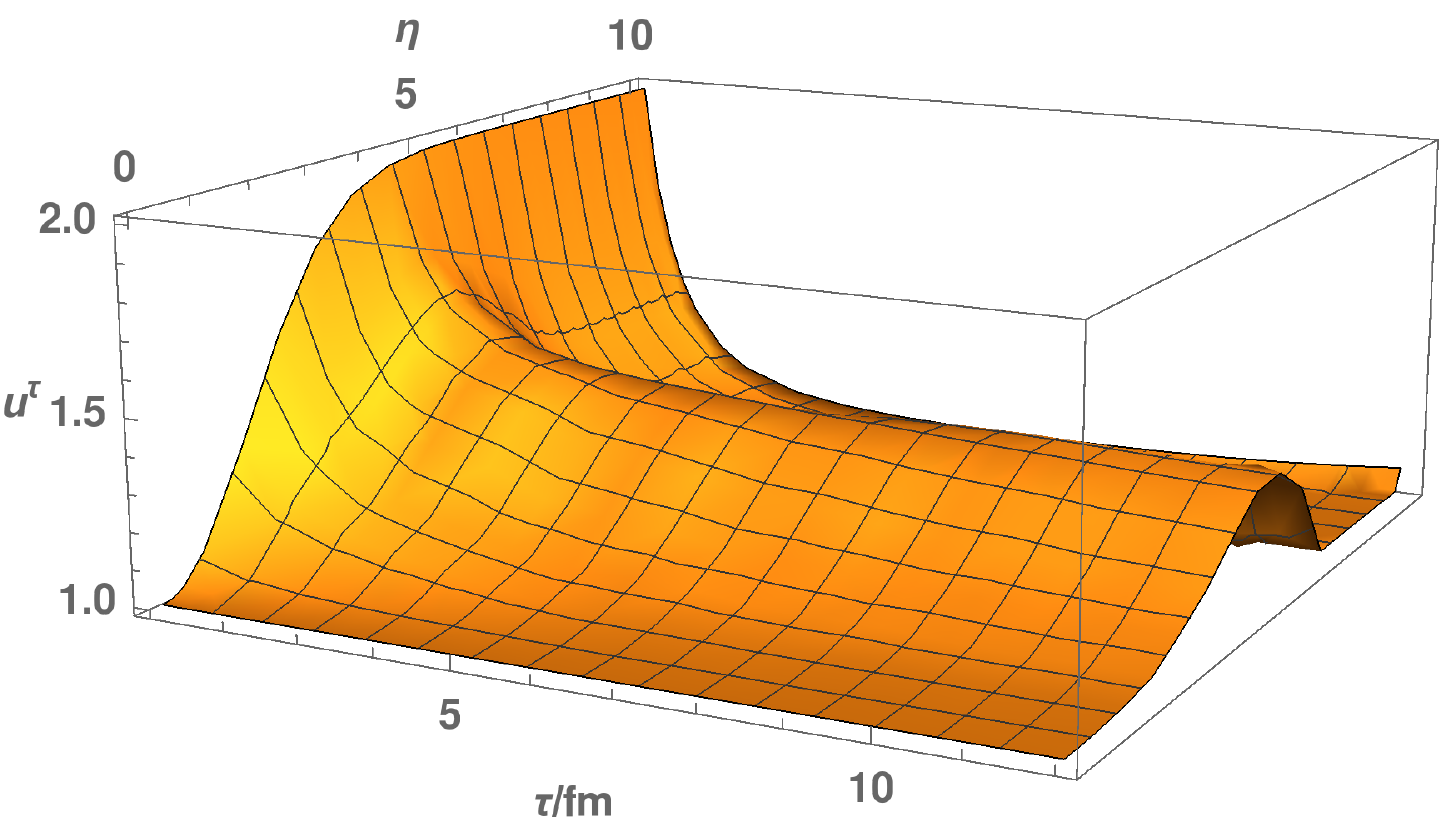}
\caption{\label{nBj240}upper left panel: $\s$ as a function of $\t$ and $\h$. $\s$ rises with oscillations as in the boost invariant case; upper right panel: $T$ as a function of $\t$ and $\h$. It shows an approximate scaling law $T\sim\t^{-\a}$ at $\h=0$; lower panel: $u^\t$ as a function of $\t$ and $\h$. From the temperature plot, the region with $\h\gsim 5$ essentially equals the $T_\mn$ on the boundary. The distinct behavior in $\h<5$ and $\h>5$ is more clearly visible in $u^\t$ plot. It is worth noting that the regions near $\h=0$ and $\h=\h_m$ have approximately Bjorken flow velocity. The region around $\h=5$ has persist extra flow at very late time. The plots correspond to $T_0=240\MeV$, $T_\mn=40\MeV$ and $u_\mx=2$.}
\end{figure}

\begin{figure}
\includegraphics[width=0.5\textwidth]{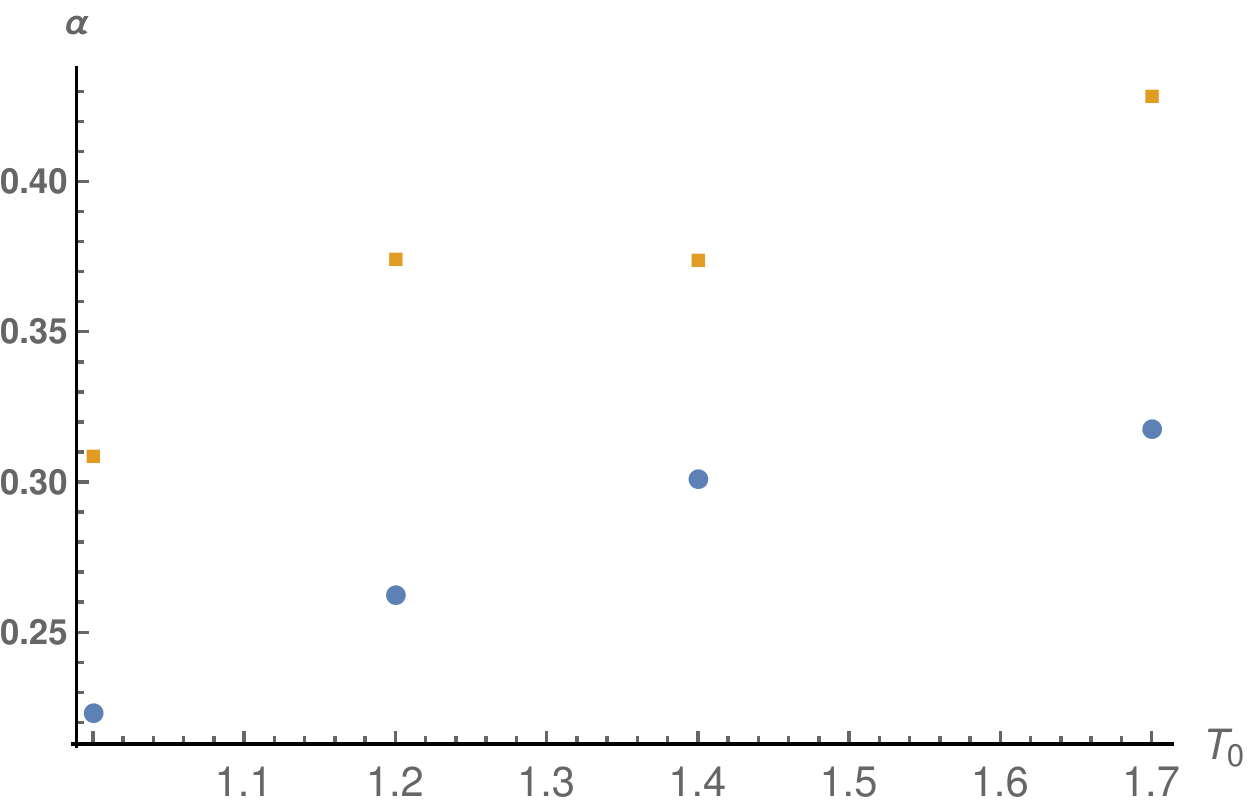}
\caption{\label{alphanBj}$\a$ as a function of $T_0$ extracted from $T(\t,\h=0)$ for non-boost invariant, together with $\a$ from boost invariant solution starting from the same $T_0$. The temperature drops quicker in non-boost invariant case, probably due to extra longitudinal flow, which takes away energy from mid-rapidity.}
\end{figure}

Now we move on to solution without boost invariance. In the absence of order parameter, Bjorken solution is supposed to describe well bulk evolution at mid-rapidity. We will see this is not quite true when order parameter is present. Without boost invariance, longitudinal flow velocity $u^\t$ becomes an independent dynamical field. We solve \eqref{eom_exp} in the interval $\h=[0,\h_m=10]$ with the following initial conditions
\begin{align}\label{ics}
&T(\t_0,\h)=(T_0-T_\mn)\(e^{-\h^2/\D^2}-e^{-\h_m/\D^2}\)+T_\mn,\no
&\s(\t_0,\h)=(\s_0-\s_\mx)\(e^{-\h^2/\D^2}-e^{-\h_m/\D^2}\)+\s_0,\no
&\pd_\t\s(\t_0,\h)=0,\no
&u^\t(\t_0,\h)=(u_\mn-u_\mx)\(e^{-\h^2/\D^2}-e^{-\h_m/\D^2}\)+u_\mx.
\end{align}
The initial profile of $T$, $\s$ and $u^\t$ are taken to be a Gaussian form with a variance $\D^2=5$. The initial maximum/minimum temperatures are at $\h=0$ and $\h=\h_m$ respectively. $\s_0$ and $\s_\mx$ are taken to be equilibrium values for the corresponding temperatures. Motivated by the Bjorken solution, we take $u_\mn=1$. Early works on longitudinal dynamics choose $u_\mx=1$ \cite{Hirano:2001eu,Satarov:2006iw}, see also \cite{Wong:2014sda,Sen:2014pfa} with Landau initial condition. In our study, keeping $u_\mx\ne1$ is necessary for the stability of solutions. We take $u_\mx$, together with $T_0$, $T_\mn$ as free parameters. With the finite interval in $\h$, we still need boundary conditions, for which we take
\begin{align}\label{bcs}
&T(\t,\h_m)=T_\mn,\no
&\s(\t,\h_m)=\s_\mx,\no
&\pd_\h\s(\t,0)=0,\no
&u^\t(\t,0)=u_\mn.
\end{align}
We impose the minimum temperature $T_\mn$ and equilibrium value of $\s$ at $\h=\h_m$. The third condition in \eqref{bcs} follows from the symmetry: $\h\leftrightarrow-\h$. We show in Fig.~\ref{nBj240} evolution of $\s$, $T$ for $T_0=240\MeV$, $T_\mn=40\MeV$ and $u_\mx=2$. Similar to the case with boost invariance, $\s$ rises with oscillations. $T$ drops with oscillations correlated with those in $\s$. As indicated in Fig.~\ref{nBj240}, the drop of temperature shows an approximate scaling law: $T\sim \t^{-\a}$ at $\h=0$.
However, the exponent $\a$ takes different values from those in boost invariant case starting with the same $T_0$, as shown in Fig.~\ref{alphanBj}. The case without boost invariance shows faster drop in temperature. This is probably caused by extra longitudinal flow, which takes away energy from the higher temperature mid-rapidity region.
The behavior of $u^\t$ is unique in the non-boost invariant case. The flow velocity near $\h=\h_m$ quickly drop to zero. The region near $\h=5$ has persistent extra longitudinal flow. While we impose $u^\t=u_\mx$ at $\h=\h_m$ at initial time, we find it quickly drop to $1$ as $\t$ increases, indicating the longitudinal flow quickly approaches Bjorken flow. At around $\h=5$, the extra longitudinal flow persists at late time.

\begin{figure}
\includegraphics[width=0.5\textwidth]{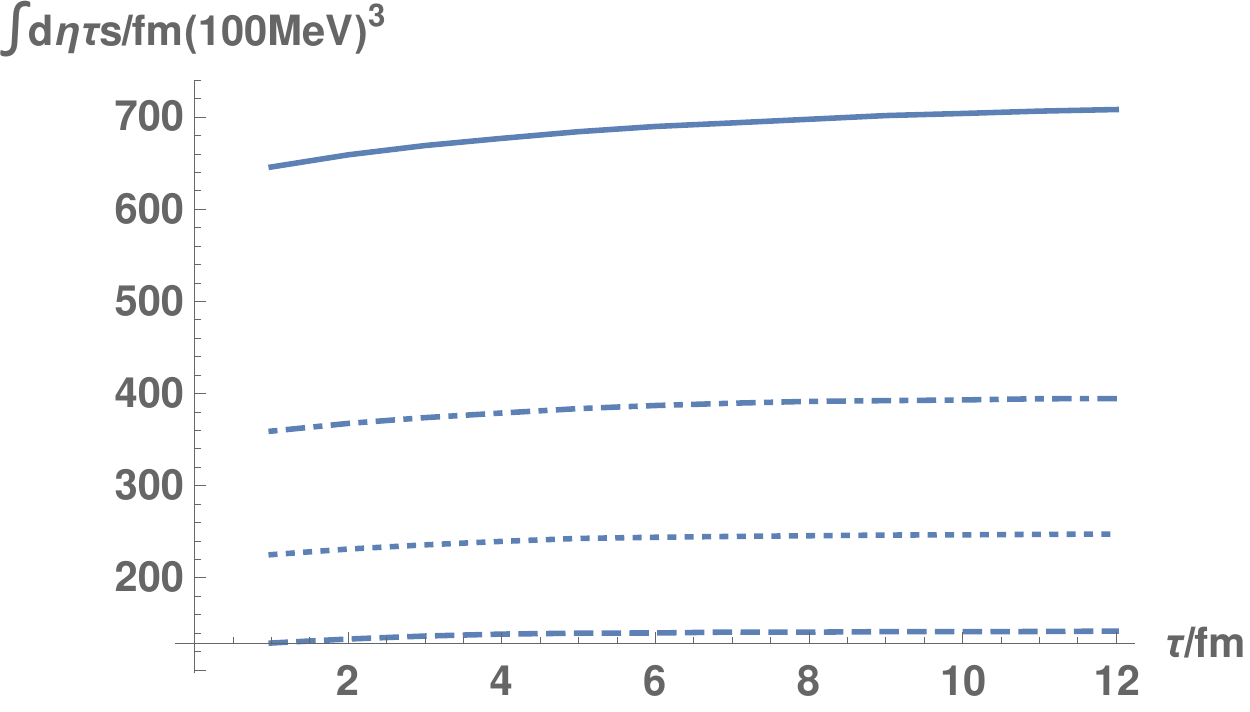}
\caption{\label{snBj}$\int d\h\t s$ as a function of $\t$ for $T_0=200\MeV$, $T_0=240\MeV$, $T_0=280\MeV$ and $T_0=340\MeV$ (bottom to top). All cases are with $u_\mx=2$. We see mild entropy production in contrast to the no entropy production in boost invariant case. We also see entropy saturates at later time for higher initial temperature.}
\end{figure}

\begin{figure}
\includegraphics[width=0.3\textwidth]{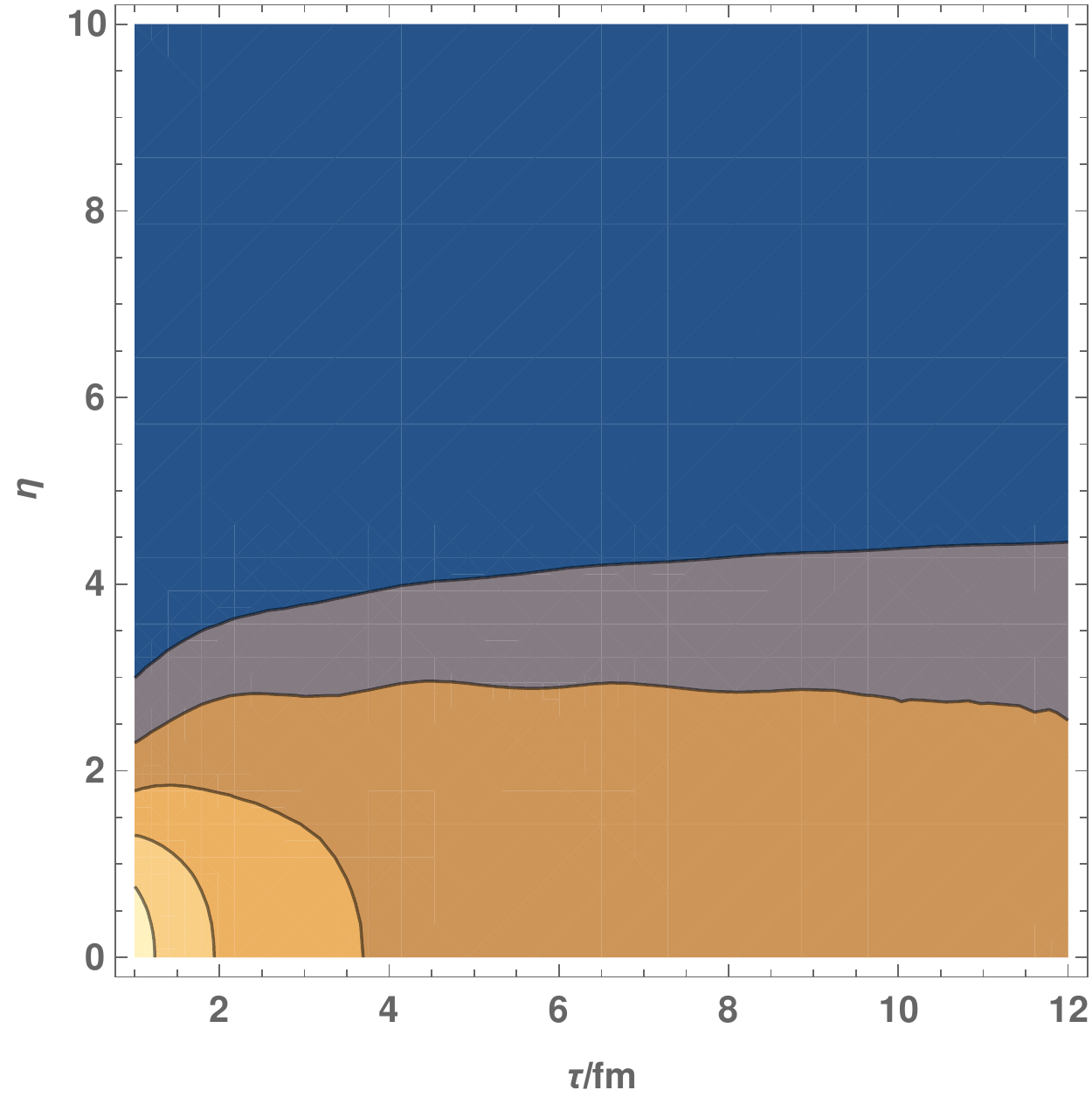}
\includegraphics[width=0.5\textwidth]{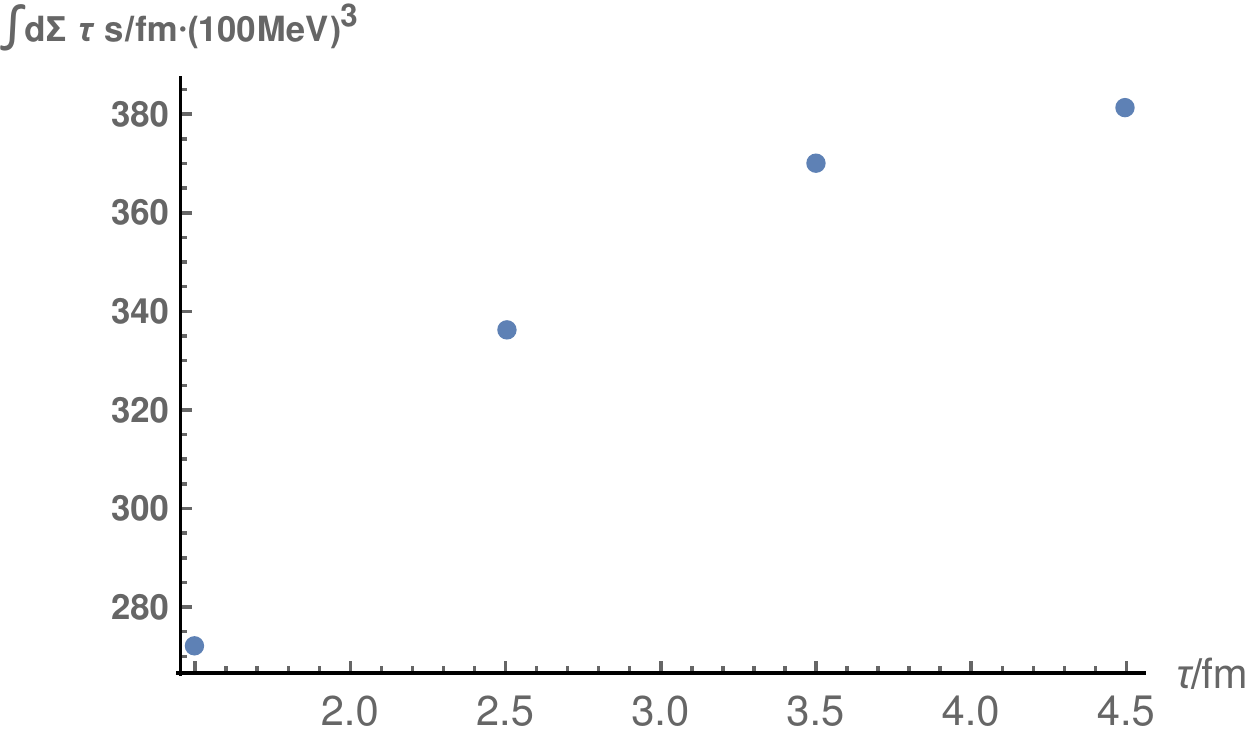}
\caption{\label{contour}left panel: isothermal contour plot for solution with $T_0=280\MeV$ and $u_\mx=2$. right panel: $\int d\Sigma\t s$ (defined in \eqref{isothermal}) as a function of contour time for $T_0=280\MeV$ and $u_\mx=2$. The contour time is chosen as the proper time at $\h=0$. We see more significant entropy production as compared to Fig.~\ref{snBj}. At late time, the entropy from constant proper time and isothermal contour agree numerically.}
\end{figure}

The behavior of entropy is qualitatively different from the boost invariant case. Integrating \eqref{Ds} with $\int\t d\t d\h d^2x_\perp$, we have
\begin{align}\label{Dsint}
0=\int\t d\t d\h D_\m\(su^\m\)=\int d\t d\h d^2x_\perp\big[\pd_\t\(\t s u^\t\)+\pd_\h\(\t s u^\h\)\big].
\end{align}
The second term on the right hand side \eqref{Dsint} is a boundary term. If we ignore it for the moment, the conserved quantity is $\t s u^\t$ rather than $\t s$, which is the entropy density per rapidity per transverse area. Therefore we expect entropy is not conserved in this case. Fig.~\ref{nBj240} suggests the rapidity averaged $u^\t$ decreases with $\t$, thus we have entropy production in non-boost invariant solution. In Fig.~\ref{snBj}, we confirm this by direct numerical integration of $\int d\h\t s$ for different initial $T_0$ and $u_\mx=2$.
The expression $\int d\h\t s$ measures total entropy per transverse area along constant proper time slice. This expression depends on our choice of time. For application to fireball evolution in heavy ion collisions, we may also use isothermal contours for integration. This definition is phenomenologically more relevant if we assume the freezeout temperature of the fireball is a constant. Fig.~\ref{contour} shows isothermal for one particular solution with $T_0=280\MeV$. On each isothermal contour, the total entropy per transverse area is given by
\begin{align}\label{isothermal}
\int d\Sigma\t s=\int \t d\h\(u^\t-u^\h\frac{\pd\t}{\pd\h}\).
\end{align}
We show total entropy per transverse area at different contour times in Fig.~\ref{contour}. A more pronounced increase with contour time is seen.
%
The increase of entropy in non-boost invariant case can be understood as follows: while the system we consider has no dissipation, it evolves starting from an off-equilibrium state, with inhomogeneity in rapidity direction. The evolution smoothens out the inhomogeneity and is irreversible. It is analogous to irreversible expansion of ideal gas in free space. This process generates net entropy. For the same reason, the boost invariant solution preserves entropy due to the absence of inhomogeneity.

\section{Conclusion and Outlook}

We studied hydrodynamics with order parameter based on linear sigma model. We obtained numerical solutions both with and without boost invariance. In both cases, we found similar behavior for the order parameter $\s$ and temperature $T$, with $\s$ rises in an oscillatory fashion. A correlated behavior is also seen in the temperature: $T$ oscillates around a power law decrease $T\sim \t^{-\a}$, as shown in Fig.~\ref{Bj200} and Fig.~\ref{nBj240}. However, the exponent for boost invariant solution is smaller than the counterpart of non-boost invariant solution with the same initial temperature at $\h=0$. This is probably due to the extra longitudinal flow in the non-boost invariant case taking away energy from the high temperature mid-rapidity region.

We also studied entropy production for the solutions. The boost invariant solution is shown to preserve entropy in the absence of dissipation. However this is no longer true for non-boost invariant solution. We used two definitions of total entropy, one from integration on constant proper time and the other on isothermal contour, with both showing entropy production. This can be understood as the evolution smoothening out the inhomogeneity of the off-equilibrium state.

While we only studied hydrodynamics at $\m_B=0$, where the phase transition is a crossover, it is more interesting to study the hydrodynamics with a first order phase transition. In the latter case, a smooth sigma profile for the initial condition would necessarily be non-monotonic, which contains larger inhomogeneity. By the inhomogeneity induced mechanism discussed in this paper, it is tempting to expect a larger entropy production. We leave more quantitative studies for the future.

\section{Acknowledgments}

S.L. is grateful to Huichao Song for useful discussions. This work is in part supported by NSFC under Grant Nos 11675274 and 11735007.

\appendix


\end{document}